\documentclass[
    ,final            % use final for the camera ready runs
  ]
  {aipproc}

\layoutstyle{6x9}

\usepackage{epsfig}

\begin{document}

\title{QCD at High Temperature : 
\\ Results from Lattice Simulations with an Imaginary $\mu$}


\author{Massimo D'Elia}{address={
  Dipartimento di Fisica, Universit\`a di Genova and INFN Sezione di Genova,
via Dodecaneso 33, I-16127 Genova, Italy}}

\author{Francesco Di Renzo}{address={Dipartimento di Fisica, 
Universit\`a di Parma and INFN, Gruppo Collegato di Parma, 
Parco Area delle Scienze 7/A, I-43100 Parma, Italy}}

\author{Maria Paola Lombardo}{address= {INFN, 
Laboratori Nazionali di Frascati, Via Enrico Fermi 40, I-00044, 
Frascati (RM), Italy}}

\begin{abstract}

We summarize our results on the phase diagram of QCD  with emphasis
on the high temperature regime. For $T \ge 1.5 T_c$ the results
are compatible with a free field behavior, while for $T \simeq 1.1 T_c$
this is not the case, clearly exposing the strongly 
interacting nature of QCD in this region.

\end{abstract}

\maketitle

%%%%%%%%%%%%%%%%%%%%%%%%%%%%%%%%%%%%%%%%%%%%
%% MAINMATTER
%%%%%%%%%%%%%%%%%%%%%%%%%%%%%%%%%%%%%%%%%%%%

\section{Introduction}

The historical developments of the phase diagram of QCD is characterized
by an increasing complication: according to early
views  based on a straightforward application of asymptotic
freedom , the phase diagram was sharply divided into an hadronic phase and 
a quark gluon plasma phase. In the late 90's it was appreciated that
the high density region is much more complicated than previously thought
\cite{Alford}. In the last couple of years, it was the turn of the 
region  above $T_c$ to become more  rich: the survival of 
bound states above the phase transition  brought up the idea of a 
more complicated, highly non--perturbative phase, whose precise nature has not
been clarified yet\cite {sQGP}. The properties of the high
temperature phase  are especially interesting
in view of the ongoing ultrarelativistic heavy ions collisions
experiments with at RHIC, which
explore temperatures close to $T_c$, and of the future experiments
at LHC, which will approach the perturbative, free gas
limit of QCD.

Up to which extent the properties of the matter produced in these
ultrarelativistic heavy ion collisions  can
be predicted by the basic theory of strong interactions, Quantum Chromo 
Dynamics? In this note we review 
our results \cite{D'Elia:2002gd,D'Elia:2004at,progress}
 on this point,

QuantumChromoDynamics (QCD) basic degrees of freedom are quarks and gluons.
The Lagrangian built by use of these fundamental fields enjoys local and
(approximate) global symmetries. The realization of the global chiral symmetry
depends on the thermodynamic conditions of the system :
it is spontaneously broken, with the accompanying phenomena of Goldstone
modes and a mass gap, in ordinary conditions, and it gets restored at high
temperature. At the same time, confinement, which is realized in the normal
phase, disappears at high temperature: all in all, in our low temperature 
world quarks are confined within hadrons, there are
light preudoscalar mesons, the Goldstone bosons, and there are
massive mesons and baryons. In the high temperature phase - the Quark Gluon 
Plasma - quarks and gluons are no longer confined, and the mass spectrum
reflects the symmetries of the Lagrangian.

In a standard nuclear physics approach these features of the two phases are
imposed by fiat, and the phase transition is obtained
by equating the free energies of the quark-gluon gas on one side, and the
hadron gas on the other side of the transition. At a variance with this,
an approach based on
QCD derives the different degrees of freedom of the two phases, as well
as the phase transition line,  from the same Lagrangian. 
These calculations, being completely non--perturbative, require a
specific technique, Lattice QCD.

\section{Lattice QCD Thermodynamics}

Without entering into the
details of this approach \cite{lat}, let us just remind that 
the QCD equations are put on a 'grid' which
should be fine enough to resolve details, and large enough to accommodate
hadrons within: obviously, this would call for grids with a large number
of points. On the other hand, the calculations complexity grows
fast with the number of nodes in the grid, and the actual choices rely
on a compromise between physics requirements and computer 
capabilities\cite{tilo}.

In practical numerical approaches, the lattice discretization is combined with
a statistical techniques for the computation of the physical observables.
This requires a positive 'measure', which is given by the exponential
of the Action. A notorious problem plagues these calculations at finite 
baryon density, as the Action itself becomes complex, with a non--positive
definite real part: because of this, for many years QCD at nonzero 
baryon density 
was not progressing at all. 

Luckily,  in the last four years a few 
lattice techniques --  imaginary chemical potential, Taylor expansion,
multiparameter reweighting -- proven successful for $\mu_B/T < 1.$
\cite{uno,due,FoPh,D'Elia:2002gd,D'Elia:2004at,tre,susc}.
It has to be stressed, however, that these techniques are just dodges 
and workaround, and do not provide a real solution to the 'sign problem'.  
Moreover, due to the computer
limitations sketched above, which we hope will be soon overcome by
the next generation of supercomputers \cite{tilo}, the results have not
yet reached the continuum, infinite volume limit.

While waiting for final results in  the scaling limit and with physical
values of the parameters, it is very useful to contrast and compare
current lattice results with model calculations
and perturbative studies.  The imaginary chemical potential 
approach\cite{Lombardo:1999cz,Hart:2000ef,FoPh,D'Elia:2002gd,D'Elia:2004at,
Giudice:2004se}to QCD thermodynamics 
seems to be ideally suited for the interpretation 
and comparison with analytic results. Results from an imaginary 
$\mu$ have been obtained for the critical line of the two, three and
two plus one flavor model \cite{FoPh}, as well as for 
four flavor \cite{D'Elia:2002gd}.
Thermodynamics results -- order parameter, pressure, number
density -- were obtained for the four flavor model
\cite{D'Elia:2004at}, and are extended in this note, where
we concentrate on the region $T > T_c$.

\begin{figure}
  \includegraphics[height=.5\textheight]{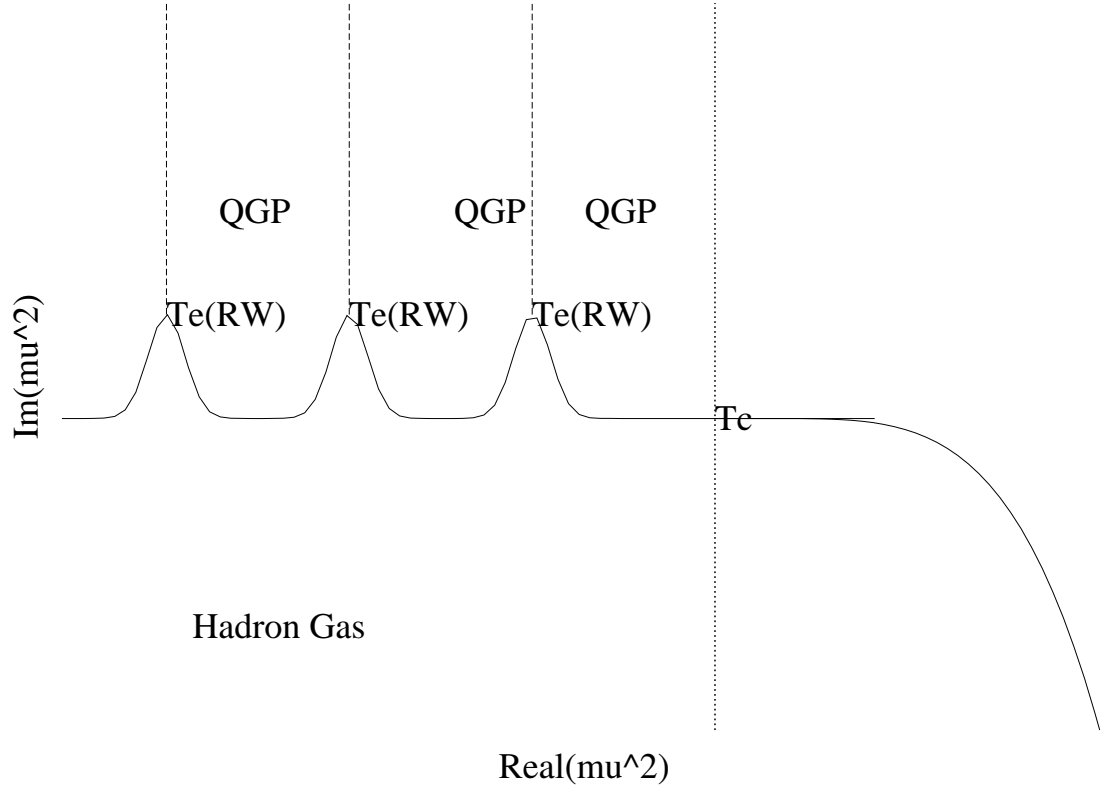}

\caption{Sketch of the phase diagram in the $\mu^2,T$ plane:
the solid line is the chiral transition, the dashed
line is the Roberge Weiss transition \cite{Roberge:1986mm}, 
the dotted line the $\mu^2=0$ axis.
 Simulations can be carried out at
$\mu^2 \le 0$ and results continued to the
physical domain $\mu^2 \ge 0$.
The derivative and reweighting methods have been used so far
to extract informations from simulations performed at $\mu=0.$
The imaginary chemical potential approaches uses results on the
left hand half plane. Different methods could be combined 
to improve the overall performance.}
\end{figure}

\section{Imaginary Chemical Potential}
The imaginary chemical potential method
 uses information from all of the negative $\mu^2$ half 
plane (Fig. 1) to explore the positive, physical relevant region.

The main physical idea behind any practical application is that 
at $\mu = 0$ fluctuations allow the exploration 
of $N_b \ne 0$ hence tell us about $\mu \ne 0$.
Mutatis mutandis, this is the same condition for the reweighting
methods to be effective: the physics of the simulation
ensemble has to overlap with that of the target ensemble.

A practical way to use the results obtained at negative
$\mu^2$ relies on their analytical continuation in the real
plane. For this to be effective\cite{Lombardo:1999cz}
${\cal Z} (\mu, T) $  must be analytical, nontrivial, and 
fulfilling:
\begin{equation}
\chi(T,\mu) = \partial \rho (\mu, T) / \partial \mu 
= \partial^2 log Z (\mu, T) / \partial \mu^2
 > 0
\end{equation}

This approach has  been tested
in the strong coupling limit \cite{Lombardo:1999cz} of QCD, in the
dimensionally reduced model of high temperature QCD
\cite{Hart:2000ef}  and, more recently,   in the two
color model \cite{Giudice:2004se}.

\section{ The Hot Phase and the approach to a Free Gas}

At high temperature, in the weak coupling regime,
finite temperature perturbation theory might serve as a guidance, 
suggesting that the first few terms of the
Taylor expansion might be adequate in
a wider range of chemical potentials.  
So, at a variance with the expansion in the hadronic phase,
where the natural parametrization is given by 
a Fourier analysis\cite{D'Elia:2002gd, D'Elia:2004at},
in this phase the natural parametrization for the
grand partition function is a polynomial. 

The leading order result for the pressure $P(T,\mu)$ in the
massless limit is easily computed, given
that at zero coupling the massless theory reduces to a non--interacting gas
of quarks and gluons, yielding for the pressure
\begin{equation}
p(T, \mu) = \frac{\pi^2}{45} T^4 \left(8 + 7 N_c \frac{n_f}{4}\right) +  \frac{n_f}{2} \mu^2 T^2
+  \frac{n_f}{4 \pi^2} \mu^4 \, .
\end{equation}
Obviously, when analytically continued to the negative
$\mu^2$ side, this gives
\begin{equation}
p(T, \mu_I) = \frac{\pi^2}{45} T^4 \left(8 + 7 N_c \frac{n_f}{4}\right) -  \frac{n_f}{2} \mu_I^2 T^2
+  \frac{n_f}{4 \pi^2} \mu_I^4 \, .
\end{equation}
Because of the Roberge Weiss \cite{Roberge:1986mm} periodicity this polynomial
behavior should be cut at the Roberge Weiss transition 
$\mu_I = \pi T / 3$:
this is consistent with the Roberge Weiss critical line
being strongly first order at high temperature.
We discuss first the results of the fits of the
number density to polynomial form; then we contrast these results
with a free field behavior. 

The considerations above suggests 
a natural ansatz for the behavior of the number density in this
phase as a simple polynomial with only odd powers. 
We performed then fits to 
\begin{equation}
n(T,\mu_I) = i a(T) \mu_I - i b(T) \mu_I^3
\end{equation}
whose obvious analytic continuation is
\begin{equation}
n(T, \mu) = a(T) \mu + b(T) \mu^3 \, .
\end{equation}
Note again that $a(T) = \chi_q(T, \mu=0)$.

In ref. \cite{D'Elia:2004at}
we contrasted the results for the particle
number at $T=1.5 T_c$, $T=2.5 T_c$, $T=3.5 T_c$  with a free
field behaviour.

Some deviations are apparent, whose origin we would like to
understand.  It would be however 
arduous, given the strong lattice artifacts, to try to make
contact with a rigorous perturbative analysis carried out in
the continuum \cite{Vuorinen:2004rd,Vuorinen:2003fs,Ipp:2003yz}.  
Rather then attempting that, we parametrize
the deviation  from a free field behavior as 
\cite{Szabo:2003kg,Letessier:2003uj}
\begin{equation}
\Delta P (T, \mu)  =  f(T, \mu) P^L_{free}(T, \mu) 
\end{equation}
where $P^L_{free}(T, \mu)$ is the lattice free result for the pressure.
For instance, in the discussion of Ref. \cite {Letessier:2003uj}
\begin{equation}
f(T, \mu) = 2(1 - 2 \alpha_s/ \pi)
\end{equation}
and the crucial point was that $\alpha_s$ is $\mu$ dependent.

We can search for such a non trivial prefactor $f(T, \mu)$ by taking 
the ratio between the numerical data and the lattice
free field result $ n^L_{free}(\mu_I)$  at imaginary chemical potential:
\begin{equation} 
R(T, \mu_I) = \frac{ n(T,  \mu_I)}{n^L_{free}( \mu_I)}
\end{equation}
A non-trivial (i.e.
not a constant) $R(T, \mu_I)$ would indicate a non-trivial 
$f(T, \mu)$.

We found  that $R(T, \mu_i)$ is constant within
errors, so that our data do not permit to distinguish a non trivial
factor within the error bars: rather, 
the results for $T \ge 1.5 T_c$ seem consistent with a free lattice
gas, with an fixed effective number of flavors $N^{eff}_f(T)/ 4 =  R(T) $:
$N^{eff}_f=  0.92 \times 4$ for $T=3.5 T_c$,  and 
$N^{eff}_f = 0.89 \times 4$ for $T = 1.5 T_c$.

One last remark concerns the mass dependence of the results, which,
as in the broken phase, can be computed from the derivative of the
chiral condensate. In the chiral limit this gives 
$\frac {\partial n}{\partial m} = 0$ , since the chiral condensate
is identically zero. We have verified that 
$\frac {\partial n}{\partial m}$ remains very small compared to
$n$ itself: in a nutshell, in the quark gluon plasma phase $<\bar \psi \psi>$
is very small (zero in the chiral limit), while the number density grows
larger, and this implies that the mass sensitivity is greatly reduced
with respect to that in the broken phase.

The discussions presented above bring  very naturally to the consideration
of a dynamical region which is comprised between the deconfinement transition,
and the endpoint of the Roberge Weiss transition.

In this dynamical region
the analytic continuation is valid till $\mu = \infty$
along the real axis, since there are no singularities for real values
of the chemical potential. 
The interval accessible to the simulations at
imaginary $\mu$ is small, as
simulations in this area hits the chiral critical
line for $\mu^2 < 0$.

 This region is of special interest and it is here that we are concentrating
our efforts: 
In Fig. 2  we show our new results 
obtained at $T/T_c = 1.095$, indicating a non trivial  deviation 
from a free field behaviour.

\begin{figure}
\includegraphics[height=.5\textheight]{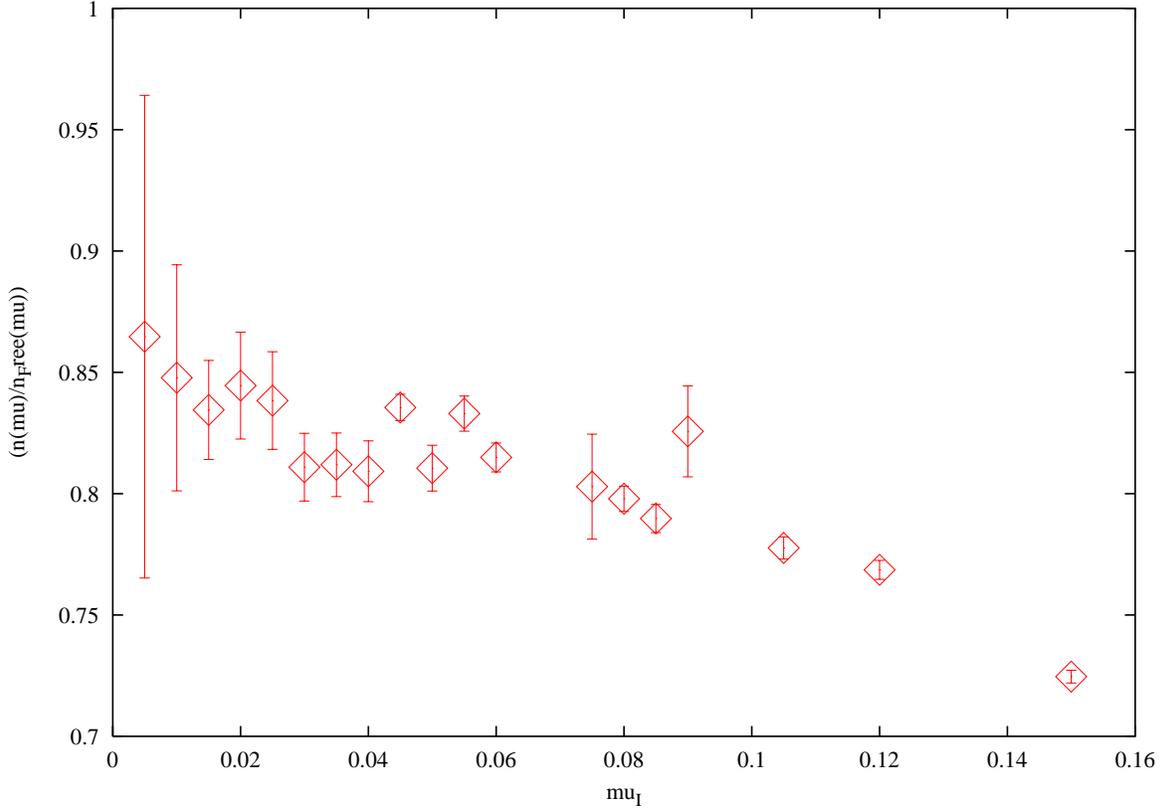}
\caption{ The ratio of the lattice results to the free
fields results $R(T, \mu_I)$ 
at $T \simeq 1.1 T_c$ : an horizontal line would indicate a nearly--free
behaviour which is clearly incompatible with the data.
 At this temperature  the chiral transition
takes place at imaginary chemical potential , i.e. at a 
negative $\mu^2$ (Fig.1)}. 
\end{figure}
Let us make some general consideration about the
thermodynamic behavior in this region 
by considering the critical line at imaginary
chemical potential. Let us consider first the case of a second
order transition: the analytic continuation
of the polynomial predicted 
by perturbation theory for positive $\mu^2$ 
would hardly reproduce the correct critical behavior
at the second order phase transition for $\mu^2 < 0$.
In fact, for a second order
chiral transition at negative $\mu^2$, 
$\Delta P (T, \mu^2) \propto (\mu^2 - {\mu^2}_c)^\chi $, 
where $\chi$ is a generic exponent. As the window between the critical
line  and the $\mu=0$ axis is anyway small, such behavior - possibly
with subcritical corrections - should persist in the proximity of the
real axis.
For generic values of the exponent a second
order chiral transition seems incompatible with a free field behavior.
The same discussion  can be repeated for a first order transition of 
finite strength, by trading the critical point $\mu_c$ with 
the spinodal point $\mu^*$ . So deviations 
from free field are to be expected in this intermediate regime.

A more detailed discussion of these results, and their interrelation
(or lack thereof)  with a strongly interactive quark gluon plasma will be 
given elsewhere \cite{progress}.

\section{Future directions}

The approach to a free gas of quarks and gluons is
a fascinating subject: we are moving from a world of colorless
hadrons to a world of colored particles - quarks, gluons, and perhaps
many more.

Three different, independent methods which
afford a quantitative
approach to this problem
have been proposed and exploited in the past few years,
producing several interesting and coherent results.
In particular we have focussed on the properties of the
hot phase right above the critical temperature, where we have
observed a clear non--perturbative behaviour for the thermodyamical
observables, showing that the system is still very far from a free
gas of quark gluons.

These results, however, still need improvement: in particular,
small quark masses,
and a controlled approach to the continuum limit. All this
 requires a large
amount of computer resources, and 
there is hope that the new dedicated supercomputers for lattice QCD -
QCDOC and apeNEXT - will produce significant advances in this field.

High quality numerical results together with a careful consideration
of phenomenological models and critical behaviour 
in the negative $\mu^2$ half--plane should produce a coherent and
complete description of the high temperature phase of the strong
interactions, which will hopefully confront soon
ongoing and future experiments.

\section*{Acknowledgments}
The new calculations reported here were performed on the APEmille
computers of the MI11 {\em Iniziativa Specifica}, and we wish
to thank our colleagues in Milano and in Parma for their kind help.

\end{document}